\documentclass[twocolumn,showpacs,preprintnumbers,amsmath,amssymb]{revtex4-1}

\usepackage{graphicx}
\usepackage{dcolumn}
\usepackage{bm}
\newcommand{\comments}[1]{} 
\usepackage{natbib}
\begin{document}


\title{Bistability in the Chemical Master Equation for Dual Phosphorylation Cycles} 

\author{Armando Bazzani}
\author{Gastone C. Castellani}
\email{Gastone.Castellani@unibo.it}
\author{Enrico Giampieri}
\author{Daniel Remondini}
\affiliation{Physics Dept. of Bologna University and INFN Bologna}%
\author{Leon N Cooper}
\affiliation{Institute for Brain and Neural Systems,Department of Physics, Brown University, Providence,
RI 02912} %

\date{\today}

\begin{abstract}
Dual phospho/dephosphorylation cycles, as well as covalent
enzymatic-catalyzed  modifications of substrates are widely diffused
within cellular systems and are crucial for the control of complex
responses such as learning, memory and cellular fate determination.
Despite the large body of deterministic studies and the increasing work
aimed at elucidating the effect of noise in such systems, some aspects
remain unclear. Here we
study the stationary distribution provided by the two-dimensional
Chemical Master
Equation for a  well-known model of a two step phospho/dephosphorylation
cycle using the quasi-steady state approximation of enzymatic
kinetics. Our aim is to
analyze the role of fluctuations and
the molecules distribution properties  in the
transition to a bistable regime. When detailed balance conditions
are satisfied
it is possible to compute equilibrium distributions  in
a closed and explicit
form. When detailed balance is not satisfied, the stationary non-equilibrium state is
strongly influenced by the
chemical fluxes. In the last case, we show how the external field derived from the generation and recombination transition rates, can be decomposed by the Helmholtz theorem, into a
conservative and a rotational (irreversible)
part. Moreover, this decomposition,  allows to compute the stationary
distribution via a perturbative approach. For a finite
number of molecules there exists diffusion dynamics in a macroscopic
region of the state space where a relevant transition rate between the  
two critical points is observed.
Further, the stationary distribution
function can be approximated by the solution of a Fokker-Planck  
equation. We illustrate the theoretical results using several numerical simulations.

\end{abstract}


\maketitle

\section{Introduction}
One of the most important aspects of biological systems is their capacity to learn and memorize patterns and to adapt themselves to exogenous and endogenous  stimuli by tuning signal transduction pathways activity. The mechanistic description of this behavior
is typically depicted as a ``switch'' that can drive the cell fate to different stable states characterized by some observables such as levels
of proteins, messengers, organelles or phenotypes \cite{xiong2003positive}. The biochemical machinery of signal transduction pathways
is largely based on enzymatic reactions, whose average kinetic can be described within the framework of chemical kinetics
and enzyme reactions as pioneered by Michaelis and
Menten \cite{segel1984modeling,michaelis1913kinetics}.
The steady state velocity equation accounts for the majority of known enzymatic reactions, and can be adjusted to the description
of regulatory properties such as cooperativity, allostericity and activation/inhibition\cite{Xie2006}. Theoretical interest in enzymatic
reactions has never stopped since Michaelis-Menten's work and has lead to new discoveries such as zero-order
ultrasensitivity \cite{goldbeter1981amplified,Bergetal2000}.
Among various enzymatic processes, a wide and important class comprises  the reversible addition and removal of phosphoric groups
via phosphorylation and dephosphorylation reactions catalyzed by kinases and phosphatases. The  phospho/dephosphorylation
cycle (PdPC) is a reversible post-translational substrate modification that is central to cellular signalling regulation and
can play  a key role in the switch phenomenon for several biological processes (\cite{Krebs1958,Samoilov2005}).
Dual PdPC's are classified as homogeneous and heterogeneous based on the number of different kinases and phosphatases (\cite{huang2009});
the homogeneous has one kinase and one phosphatase, while the heterogeneous has two kinases and two phosphatases. Variants of homogeneous
and heterogeneous dual PdPC's may only have  a non-specific phosphatase and two specific kinases \cite{Cast01} or, symmetrically,
a non-specific kinase and two specific phosphatases.
The PdPC  with the non-specific phosphatase controls the phosphorylation state of AMPA receptors that mediates induction
of Long Term Potentiation (LTP) and Long Term Depression (LTD) in vitro and in vivo\cite{Cast01, Cast09,Bear2006}.
Recently, several authors have reported bistability in homogeneus pPdPC \cite{ortega2006,huang2009}) as well as those with a
non-specific phosphatase and two different kinases \cite{Cast09}.
The bistable behaviour of the homogeneous system is explained on the basis of a competition between the substrates for the enzymes.
The majority of studies on biophysical analysis of phospho/dephosphorylation cycle have been performed in a averaged, deterministic framework based on
Michaelis-Menten (MM) approach, using the steady state approximation. However, recently some authors \cite{Mettet07} have pointed to the role of
fluctuations in the dynamics of biochemical reactions.
Indeed, in a single cell, the concentration of molecules (substrates and enzymes) can be low, and thus it is necessary to study the 
PdPC cycle within a stochastic framework. A ``natural'' way to cope with this problem is the
so-called Chemical Master Equation (CME) approach \cite{vanKampen}, that realizes in an exact way the probabilistic dynamics of a finite
number of molecules, and recovers the chemical kinetics of the Law of Mass Action, which yields the continuous Michaelis-Menten equation in the thermodynamic limit ($N\rightarrow \infty$,)
using the mean field approximation. In this paper we study a stochastic formulation of enzymatic cycles that
has been extensively considered by several authors (\cite{huang2009,ortega2006}). The deterministic descriptions of these models characterize the stability
of fixed points and give a geometrical interpretation of the observed steady states, as the intersection of conic curves(\cite{ortega2006}).
The stochastic description can in fact  provide further information
on the relative stability of the different steady states in
terms of a stationary distribution.
We propose a perturbative approach for computing the stationary solution out of the thermodynamics equilibrium. 
 We also point out the role of
currents in the transition from a mono-modal distribution to a bimodal distribution; this corresponds to bifurcation 
in the deterministic approach. The possibility that chemical fluxes control the distribution shape suggests a generic mechanism
used by biochemical systems out of thermodynamic equilibrium to obtain a plastic behavior.
Moreover, we show that at the bimodal transition there exists a diffusion region in the configuration space where a Fokker-Planck
equation can be introduced to approximate the stationary solution.
Analogous models have been previously studied \cite{Bergetal2000,qian2003thermodynamic,qian2008temporal} for single-step PdPC.

\section{Dual phosporylation/dephosphorylation enzymatic cycles}

\begin{figure}[th]
\centerline{\includegraphics[width=0.5\textwidth]{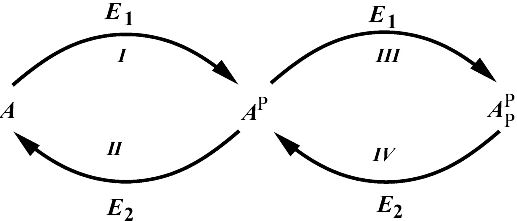}} 
\caption[]{Scheme of the double enzymatic cycle of addition/removal reactions of chemical groups via  Michelis-Menten kinetic equations as shown in eq (\ref{double_cycle_eq_two_enz}) in the case of phosphoric groups.}
\label{double_cycle_gen}
\end{figure}

The process shown in  Fig. (\ref{double_cycle_gen})
is a two-step chain  of addition/removal reactions of chemical groups and may, in general,
model important biological processes such as phenotype switching (ultrasensitivity) and chromatine
modification by histone acetylation/deacetylation as well as phospho/dephosphorylation reactions. Without loss of generality, we perform a detailed study of the homogeneous  phospho/dephosphorylation two-step cycles (PdPC cycles) where two enzymes drive phosphorylation and dephosphorylation respectively. Thus,  there is a competition between the two cycles for the advancement of the respective reactions.

The deterministic Michaelis-Menten (MM) equations of the scheme (\ref{double_cycle_gen}) with the quasi-steady-state hypothesis reads:


\begin{eqnarray}
\frac{dA}{dt}&=&\frac{k_{M_4}v_{M_2} (1-A-A^P_P)}{k_{M_2}k_{M_4} +k_{M_4}( 1-A-A^P_P)+k_{M_2}A^P_P}\nonumber \\ 
& -&\frac{k_{\tiny{M}_3}v_{M_1} A}{k_{M_1}k_{M_3}+k_{M_1}(1-A-A^P_P)+k_{M_3}A}\nonumber \\
\frac{d A^P_P}{dt} &=& \frac{k_{M_1}v_{M_3}(1-A-A^P_P)}{k_{M_1}k_{M_3} +k_{M_1}(1-A-A^P_P)+k_{M_3}A}\nonumber \\
& -&\frac{k_{M_2}v_{M_4} A^P_P}{k_{M_2}k_{M_4} +k_{M_4}( 1-A-A^P_P)+k_{M_2}A^P_P } 
\label{double_cycle_eq_two_enz}
\end{eqnarray}
where $A$ and $A^P_P$ are the concentrations of the non-phosphorylated and double phosphorylated substrates, $k_{M_{i}}$ denote the MM constants and $v_{M_{i}}$ are the maximal reaction velocities ($i=1,\cdots 4$). Let $n_1$ and $n_2$ denote the molecules number of the substrates $A$ and $A_P^P$ respectively,
the corresponding CME for the probability distribution $\rho(n_1,n_2,t)$ is written
\begin{eqnarray}
{\partial\rho\over \partial t}&=&g_1(n_1-1,n_2)\rho(n_1-1,n_2,t)-g_1(n_1,n_2)\rho(n_1,n_2,t)\nonumber\\
&+&r_1(n_1+1,n_2)\rho(n_1+1,n_2,t)-r_1(n_1,n_2)\rho(n_1,n_2,t)\nonumber\\
&+&g_2(n_1,n_2-1)\rho(n_1,n_2-1,t)-g_2(n_1,n_2)\rho(n_1,n_2,t)\nonumber\\
&+&r_2(n_1,n_2+1)\rho(n_1,n_2+1,t)-r_2(n_1,n_2)\rho(n_1,n_2,t)\nonumber\\
\label{markov1s}
\end{eqnarray}
where  $g_j(n_1,n_2)$ and $r_j(n_1,n_2)$ are the generation and recombination terms respectively, defined as :
\begin{eqnarray}
r_1(n_1,n_2)=\frac{K_{M_3}v_{M_1}n_1}{K_{M_1}K_{M_3} +K_{M_1}(N_T-n_1-n_2)+K_{M_3}n_1}\nonumber\\
g_1(n_1,n_2)=\frac{K_{M_4}v_{M_2}(N_T-n_1-n_2)}{K_{M_2}K_{M_4} + K_{M_4}(N_T-n_1-n_2)+K_{M_2}n_2} \nonumber\\
r_2(n_1,n_2)=\frac{K_{M_2}v_{M_4}n_2}{K_{M_2}K_{M_4} + K_{M_4}(N_T-n_1-n_2)+K_{M_2}n_2}\nonumber\\
g_2(n_1,n_2)=\frac{K_{M_1}v_{M_3}(N_T-n_1-n_2)}{K_{M_1}K_{M_3} +K_{M_1}(N_T-n_1-n_2)+K_{M_3} n_1} \nonumber\\
\label{twoenz}
\end{eqnarray}
$N_T$ is the total number of molecules, and we have introduced scaled constants $K_M=N_T k_M$. The
biochemical meaning is that the enzyme quantities should scale as the total number of molecule $N_T$, to have
a finite thermodynamic limit $N_T\to\infty$, in the transition rates (\ref{twoenz}).
As it is known from the theory of one-step Markov processes, the CME (\ref{markov1s})
has a unique stationary solution $\rho_s(n_1,n_2)$ that describes the statistical properties of the system on a long time scale.
The CME recovers the Mass Action-based MM equation (\ref{double_cycle_eq_two_enz}) in the thermodynamic limit when the average field theory
approach applies. Indeed it can be shown that the critical points of the stationary distribution
for the CME can be approximately computed by the conditions (cfr. eq. (\ref{crit_point}))
\begin{equation}
g_1(n_1,n_2) = r_1(n_1+1,n_2)\quad
g_2(n_1,n_2)= r_2(n_1,n_2+1)
\label{fixedp}
\end{equation}
whose solutions tend to the equilibrium points of the MM equation when the fluctuation effects are reduced in the thermodynamic limit
as $O(1/\sqrt{N})$. As a consequence, one would expect that the probability distribution becomes singular, being concentrated at the fixed stable points of the equations (\ref{double_cycle_eq_two_enz}), and that the transition rate among the
stability regions of attractive points is negligible. However, when a phase transition occurs due to the bifurcation of the stable solution,
fluctuations are relevant even for large $N_T$ and the CME approach is necessary.
In the next section we discuss the stationary distribution properties for the CME (\ref{markov1s}).

\section{The Stationary Distribution }

The stationary solution $\rho_s(n_1,n_2)$ of the CME (\ref{markov1s}) can be characterized by a discrete version
of the zero divergence condition for the current vector $\vec J$ components(see Appendix)
\begin{eqnarray}
J_1^s&=& g_1(n_1-1,n_2)\rho_s(n_1-1,n_2)-r_1(n_1,n_2)\rho_s(n_1,n_2)\nonumber \\
J_2^s&=& g_2(n_1,n_2-1)\rho_s(n_1,n_2-1)-r_2(n_1,n_2)\rho_s(n_1,n_2)\nonumber \\
\label{current}
\end{eqnarray}
and the CME r.h.s.  reads
\begin{equation}
D_1 J_1^s(n_1,n_2)+D_2 J_2^s(n_1,n_2)=0
\label{staz_sol}
\end{equation}
where we have introduced the difference operators
\begin{eqnarray}
D_1 f(n_1,n_2)=f(n_1+1,n_2)-f(n_1,n_2)\nonumber\\
D_2 f(n_1,n_2)=f(n_1,n_2+1)-f(n_1,n_2)
\label{diff_op}
\end{eqnarray}
Due to the commutative properties of the difference operators, the zero-divergence condition for the current is
equivalent to the existence of a current potential $A(n_1,n_2)$ such that
\begin{eqnarray}
J_1^s(n_1,n_2)&=&  D_2 A(n_1,n_2) \quad n_1\ge 1 \nonumber\\
J_2^s(n_1,n_2)&=& -D_1 A(n_1,n_2) \quad n_2\ge 1 \nonumber\\
\label{rotore}
\end{eqnarray}
We remark that the r.h.s. of eq. (\ref{rotore}) is a discrete version of the curl operator.
The potential difference $A(n_1',n_2')-A(n_1,n_2)$ defines the chemical transport across any line connecting the states $(n_1,n_2)$ and $(n_1',n_2')$.
At the stationary state the net transport across any closed path is zero and we have no current source in the network.
As discussed in \cite{vanKampen,Schnak76} we distinguish two cases: when the potential $A(n_1,n_2)$ is constant (the so called ``detailed balance condition") and
the converse case corresponding to a non-equilibrium stationary state.
In this case the stationary solution $\rho_s$ is characterized by the condition $J_1=J_2=0$
over all the states, whereas in the other case we have macroscopic chemical fluxes in the system.
When detailed balance holds, simple algebraic manipulations (see Appendix) result in the following conditions for the stationary solutions
\begin{eqnarray}
D_1\ln\rho_s(n_1,n_2)&=&\ln a_1(n_1,n_2)\\
D_2\ln\rho_s(n_1,n_2)&=&\ln a_2(n_1,n_2) \qquad n_1+n_2<N_T \nonumber \\
\label{det_bal_staz}
\end{eqnarray}
where the discrete drift vector field components $a_i$ are defined by:
\begin{equation}
a_1(n_1,n_2)={g_1(n_1,n_2)\over r_1(n_1+1,n_2)}\quad
a_2(n_1,n_2)={g_2(n_1,n_2)\over r_2(n_1,n_2+1)}.
\label{field}
\end{equation}
Equations (9) and (\ref{det_bal_staz}) imply the existence of a potential $V(n_1,n_2)$ such that
\begin{eqnarray}
\ln a_1(n_1,n_2)&=&-D_1 V(n_1,n_2)\nonumber \\
\ln a_2(n_1,n_2)&=&-D_2 V(n_1,n_2)
\label{det_bal2}
\end{eqnarray}

\begin{figure*}[th!]
\includegraphics[width=0.45\textwidth] {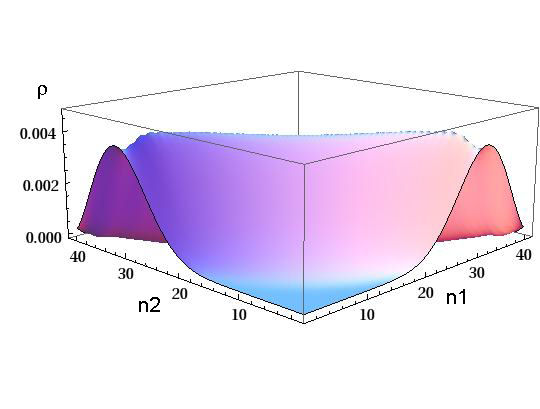} 
\qquad
\includegraphics[width=0.45\textwidth] {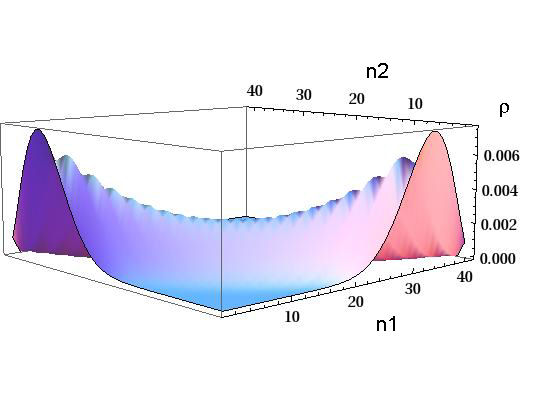} 
\caption[]{Stationary distributions for the $A$ and $A_P^P$ states in the double phosphorylation cycle when detailed balance
(\ref{det_bal_cond}) holds with $K_{M_1}=K_{M_4}=1$ and $K_{M_2}=K_{M_3}=2$. In the top figure we set the reaction velocities
$v_{M_1}=v_{M_2}=1$ and $v_{M_2}=v_{M_3}=1.05$ (symmetric case), whereas in the bottom
figure we increase the $v_{M_2}$ and $v_{M_3}$ value to 1.15. The number of molecules is $N_T=40$.
The transition from a unimodal distribution to a bimodal distribution is clearly visible.}
\label{detbal_fig}
\end{figure*}

Using definition (\ref{twoenz}) it is possible to explicitly compute a set of parameter values for the PdPC cycle,
that satisfy the detailed balance condition (\ref{det_bal2}) according to the relations
\begin{equation}
2K_{M_1}=K_{M_2}=K_{M_3}=2K_{M_4}
\label{det_bal_cond}
\end{equation}
where the reaction velocities $V_M$ are arbitrary. The stationary distribution is given by the Maxwell-Boltzmann distribution
\begin{equation}
\rho_s(n_1,n_2)=\exp(-V(n_1,n_2))
\label{max_boltz}
\end{equation}
where the potential $V(n_1,n_2)$ is computed by integrating equation (\ref{det_bal2}) and choosing the initial value $V(0,0)$
to normalize the distribution (\ref{max_boltz}). Using a thermodynamical analogy, we can interpret the  potential difference $V(n_1,n_2)-V(0,0)$
as the chemical energy needed to reach the state $(n_1,n_2)$ from the initial state $(0,0)$. As a consequence, the vector field (\ref{field})
represents the work for one-step transition along the $n_1$ or $n_2$ direction. Definition (\ref{max_boltz}) also implies that the critical
points of the stationary distribution are characterized by the conditions
\begin{equation}
{g_1(n_1,n_2)\over r_1(n_1+1,n_2)}={g_2(n_1,n_2)\over r_2(n_1,n_2+1)}=1
\label{crit_point}
\end{equation}
and coincides with the critical points of the MM equations.  In figure \ref{detbal_fig} we plot the stationary distributions (\ref{max_boltz})
in the case $N_T=40$ with $K_{M_1}=K_{M_4}=1$ and $K_{M_2}=K_{M_3}=2$; we consider two symmetric cases: $v_{M_1}=v_{M_2}=1$ with $v_{M_2}=v_{M_3}=1.05$
or $v_{M_2}=v_{M_3}=1.15$ (all the units are arbitrary). In the first case, the probability distribution is unimodal, whereas in the second case
the transition to a bimodal distribution is observed. Indeed, the system has a phase transition at $v_{M_2}=v_{M_3}\simeq 1.1$
that corresponds to a bifurcation of the critical point defined by the condition (\ref{crit_point}).

In figure \ref{detbal_fig} we distinguish two regions: a drift dominated region and a diffusion dominated region.
In the first region the chemical reactions mainly follow the gradient of the potential $V(n_1,n_2)$ and tend to concentrate around the stable critical points, so that
the dynamic is well described by a Liouville equation\cite{Huang87}.
In the second region the drift field (\ref{field}) is small and the fluctuations due to the finite size introduce a diffusive behaviour. Then
the distribution can be approximated by the solution of a Fokker-Planck equation\cite{vanKampen}.
As discussed in the Appendix, the diffusion region is approximately determined by the conditions
\begin{eqnarray}
 g_1(n_1,n_2)-r_1(n_1+1,n_2)&\simeq& g_2(n_1,n_2)-r_2(n_1,n_2+1)\nonumber\\
&\simeq & O(1/N_T)
\label{con_cond}
\end{eqnarray}

\begin{figure}[th]
\centerline{\includegraphics[width=0.45\textwidth]{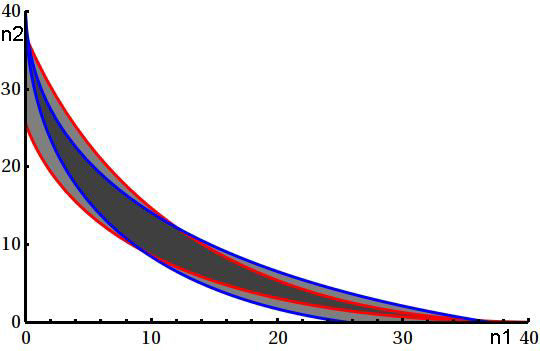}} 
\caption[]{In grey we show the region where components of the vector field (\ref{field}) are $\simeq 1$ using the parameter values of fig. \ref{detbal_fig} (bottom). The blue lines enclose the region where the first component is nearby 1, whereas the red ones enclose the corresponding region for the second component.}
\label{ellissi_fig}
\end{figure}

To illustrate this phenomenon, we outline in fig. \ref{ellissi_fig} the region where condition (\ref{con_cond}) is satisfied (i.e. the gradient of the potential $V(n_1,n_2)$ is close to 0 (\ref{det_bal2})). This is the region
where the fixed points of the MM equation are located, and comparison with fig. \ref{detbal_fig} shows that it defines the support of
the stationary distribution.

In the diffusion dominated region a molecule can undergo a transition from the dephosphorylated equilibrium to the
double phosphorylated one. At the stationary state the transition probability from one equilibrium to the other can be estimated by the Fokker-Planck
approximation; but this does not imply that the Fokker-Planck equation allows us to describe the transient relaxation process toward the stationary
state. Indeed, due to the singularity of the thermodynamics limit, the dynamics of transient states may depend critically on
finite size effects not described by using the Fokker-Planck approximation\cite{Vellela2007}.
To cope with these problems in the PdPC model further studies are necessary.\par\noindent
When the current (\ref{rotore}) is not zero the CME (\ref{markov1s}) relaxes toward a Non-Equilibrium Stationary State (NESS)
and the field (\ref{field}) is not conservative. We shall perform a perturbative approach to point out the effects of currents
on the NESS nearby the transition region to the bimodal regime. We consider the following decomposition for the vector field
\begin{eqnarray}
\ln a_1(n_1,n_2)&=&-D_1 V_0(n_1,n_2)+D_2 H(n_1,n_2) \nonumber\\
\ln a_2(n_1,n_2)&=&-D_2 V_0(n_1,n_2)-D_1 H(n_1,n_2) \nonumber\\
&\quad & n_1+n_2\le N_T-1 
\label{field_nc}
\end{eqnarray}
where the rotor potential $H(n_1,n_2)$ takes into account the irreversible rotational part. The potential
$H(n_1,n_2)$ can be recursively computed using the discrete Laplace equation
\begin{equation}
(D_1 D_1 +D_2D_2)H=D_2\ln\left (a_1(n_1,n_2)\right )-
D_1\ln\left (a_2(n_1,n_2)\right )
\label{laplace}
\end{equation}
where $n_1+n_2\le N_T-2$, with the boundary conditions $H(n,N-n)=H(n,N-1-n)=0$ (see Appendix). Assuming that the potential $H$ is small with respect to $V$, we can
approximate the NESS by using the Maxwell-Boltzmann distribution (\ref{max_boltz}) with $V=V_0$. However,
as we shall show in the next section, at the phase transition even the effect of small currents becomes critical, and the study
of higher perturbative orders is necessary. To point out the relation among the rotor potential $H$, the
NESS $\rho_s$ and the chemical flux $J$, we compute the first perturbative order letting  $\rho_s(n_1,n_2)=\exp(-V_0(n_1,n_2)-V_1(n_1,n_2))$.
From definition (\ref{current}) the condition (8) reads:
\begin{eqnarray}
&&\exp\left (-V_0(n_1,n_2)-V_1(n_1,n_2)\right ) r_1(n_1,n_2) \nonumber\\
&&\cdot\left (\exp\left( D_2 H(n_1-1,n_2)\right )-\exp\left (-D_1V_1(n_1-1,n_2)\right )\right ) \nonumber\\
&&= D_2 a(n_1,n_2))\nonumber\\
&&\exp\left (-V_0(n_1,n_2)-V_1(n_1,n_2)\right ) r_2(n_1,n_2) \nonumber\\
&&\cdot\left (\exp\left (-D_1 H(n_1,n_2-1)\right )-\exp\left (-D_2 V_1(n_1,n_2-1)\right )\right ) \nonumber\\
&&=-D_1 a(n_1,n_2)  \nonumber\\
\label{current_nq}
\end{eqnarray}

$V_1(n_1,n_2)$ turns out to be an effective potential that simulates the current's effect on the
unperturbed stationary distribution by using a conservative force.
We note that if the rotor potential $H$ is zero, then both the current potential $A$ and the potential correction $V_1$ are zero, so that
all these quantities are of the same perturbative order, and the first perturbative order of  eqs. (\ref{current_nq}) reads
\begin{eqnarray}
&&\exp\left (-V_0(n_1,n_2)\right ) r_1(n_1,n_2)\cdot\nonumber\\
&&\left (D_1 V_1(n_1-1,n_2)+D_2 H(n_1-1,n_2)\right )=D_2 A(n_1,n_2)  \nonumber\\
&&\exp\left (-V_0(n_1,n_2)\right ) r_2(n_1,n_2)\cdot\nonumber\\
&&\left (D_2 V_1(n_1,n_2-1)-D_1 H(n_1,n_2-1)\right )=-D_1 A(n_1,n_2)\nonumber\\
\label{current_nq1}
\end{eqnarray}
From the previous equations we see that the currents depend both on the
rotor potential $H$ and the potential correction $V_1$, which is
unknown; thus they cannot be directly computed from (\ref{current_nq1}). We obtain an equation
for $V_1$ by eliminating the potential $A$ from (\ref{current_nq1})
\begin{eqnarray}
&&D_1\left (\exp\left (-V_0(n_1,n_2)\right ) r_1(n_1,n_2)D_1 V_1(n_1-1,n_2)\right )\nonumber \\
&&+D_2\left (\exp\left (-V_0(n_1,n_2)\right ) r_2(n_1,n_2)D_2 V_1(n_1,n_2-1)\right )=\nonumber \\
&&D_2\left (\exp\left (-V_0(n_1,n_2)\right ) r_2(n_1,n_2)D_1 H(n_1,n_2-1) \right )\nonumber \\
&&-D_1\left (\exp\left (-V_0(n_1,n_2)\right )r_1(n_1,n_2)D_2 H(n_1-1,n_2)\right )\nonumber \\
\label{pot_pert}
\end{eqnarray}
Eq. (\ref{pot_pert}) is defined for $n_1\ge 1$, $n_2\ge 1$ and $n_1+n_2\le N_T-1$, and we can solve the system by introducing the boundary conditions
$V_1(n,0)=V_1(0,n)=V_1(n,N_T-n)=0$. It is interesting to analyze equation (\ref{pot_pert}) in the phase transition regime.
When the recombination terms $r_1$,$r_2$ are almost equal and their variation is small (for our parameter choice this is true for $n_1\simeq n_2$
and $n_1+n_2\gg 1$) the r.h.s. can be approximated by
\begin{eqnarray}
\exp(-V_0(n_1,n_2)\left [D_2 V_0(n_1,n_2) D_1 H(n_1,n_2-1)\right.\nonumber \\
\left.\qquad-D_1 V_0(n_1,n_2)D_2 H(n_1-1,n_2) \right ]\nonumber \\
\end{eqnarray}
As a consequence, in the transition regime this term is negligible since both $D_1 V_0(n_1,n_2)$ and $D_2 V_0(n_1,n_2)$ tend toward zero in the diffusion region
where the bifurcation occurs; thus the first perturbative order is not enough to compute the stationary distribution correction, but higher
orders should be considered.

\begin{figure*}[th]
 \centerline{
\includegraphics[width=0.4\textwidth]{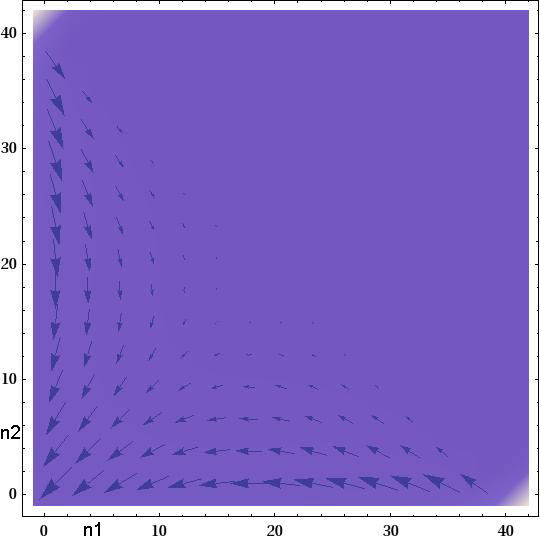}
\qquad
\includegraphics[width=0.4\textwidth]{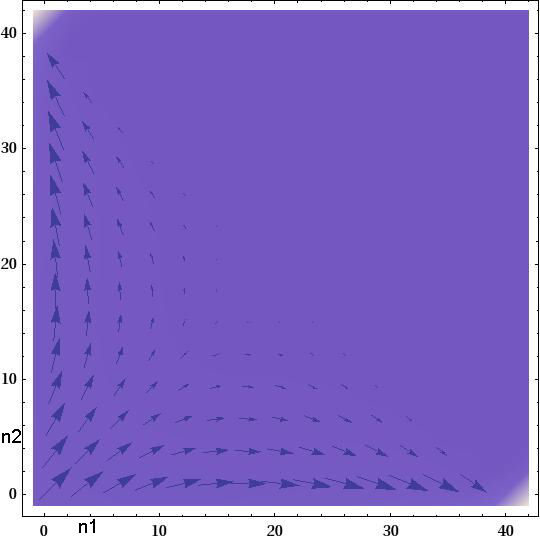}
}
\caption[]{Plot of the rotor field for the potential $H$ using the following parameter values: $v_{M1}=v_{M2}=1$, $v_{M_2}=v_{M_3}=1.15$ $K_{M_1}=K_{M_4}=1$, $N_T=40$ and $K_{M_2}=K_{M_3}=1.8$ (left picture) or $K_{M_2}=K_{M_3}=2.2$ (right picture).}
\label{rot_fig}
\end{figure*}

\begin{figure*}[ht]
\centerline{
\includegraphics[width=0.45\textwidth]{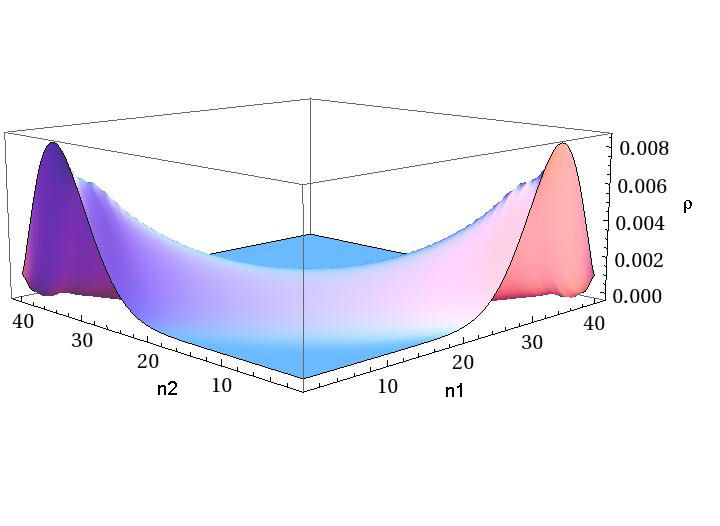}
\qquad
\includegraphics[width=0.45\textwidth]{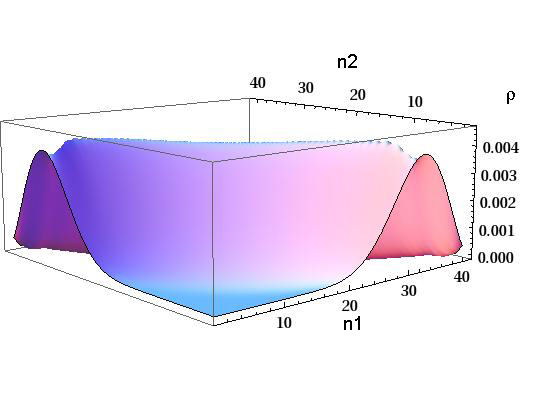}
}
\caption[]{(Left picture): plot of the zero-order approximation for the probability distribution using the decomposition (\ref{field_nc}) for the
vector field associated to the CME. (Right picture): plot of the stationary distribution computed by directly solving the CME
(\ref{markov1s}). We use the parameter values of the case I in the table \ref{table1}.}
\label{dist01_fig}
\end{figure*}

\begin{figure*}[ht]
\centerline{
\includegraphics[width=0.45\textwidth]{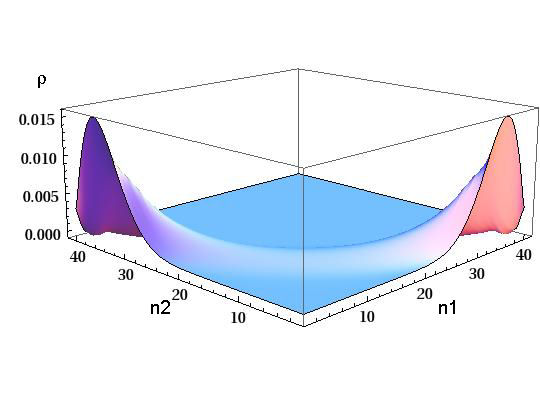}
\qquad
\includegraphics[width=0.45\textwidth]{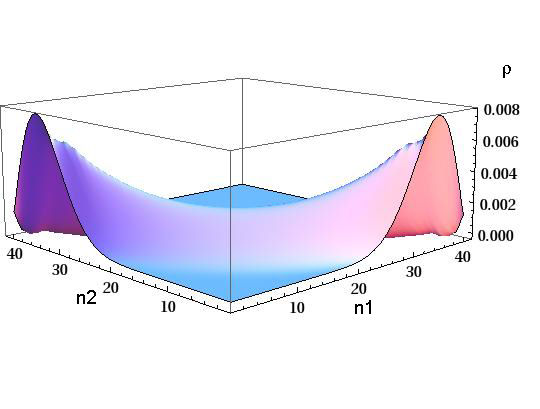}
}
\caption[]{The same as in fig. \ref{dist01_fig} using parameter values of case II in table \ref{table1}.}
\label{dist02_fig}
\end{figure*}

\begin{figure*}[ht]
\centerline{
\includegraphics[width=0.45\textwidth]{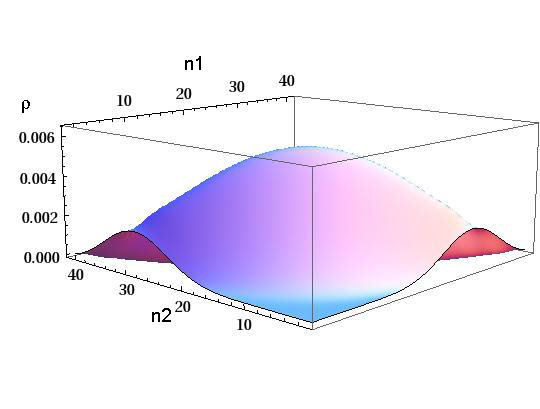}
\qquad
\includegraphics[width=0.45\textwidth]{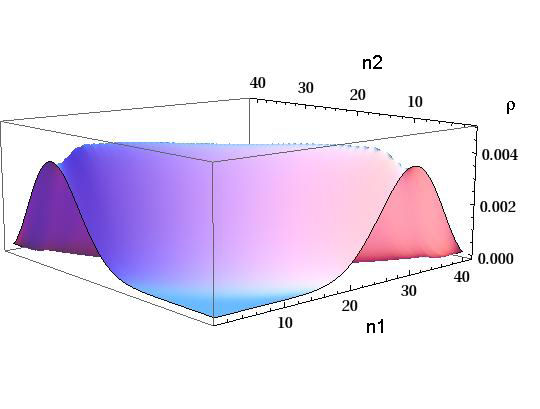}
}
\caption[]{The same as in fig. \ref{dist01_fig} using the parameter values of the case III in table \ref{table1}.}
\label{dist03_fig}
\end{figure*}

\section{Numerical simulations}

In order to study the non equilibrium stationary conditions in the double phosphorylation cycle (\ref{double_cycle_eq_two_enz})
we have perturbed the detailed balance conditions considered in figure (\ref{detbal_fig}) by changing the value of the MM constants $K_{M_2}$
and $K_{M_3}$. In figure (\ref{rot_fig}) we show the rotor field of the potential $H$ in the cases $K_{M_2}=K_{M_3}=2.2$ and $K_{M_2}=K_{M_3}=1.8$
when the detailed balance condition (\ref{det_bal_cond}) does not hold.
We see that in the first case (left picture) the rotor field tends to move the particles from the borders toward the central region $n_1=n_2$,
so that we expect an increase of the probability distribution in the center, whereas in the second case the rotor field is directed
from the central region to the borders and we expect a decrease of the probability distribution in this region. To illustrate the effect of the potential $H$,
we compare the zero-order approximation of the probability distribution (\ref{max_boltz}), where the potential $V_0(n_1,n_2)$ is computed  using decomposition
(\ref{field_nc}) with the stationary solution of the CME (\ref{markov1s}). The main
parameter values are reported in table \ref{table1}
\begin{center}
\begin{tabular}{l*{6}{c}r}
Case   & $K_{M_2}$ & $K_{M_3}$  & $v_{M_2}$ & $v_{M_3}$ \\
\hline
I  & 1.8 & 1.8 & 1.05 & 1.05 \\
II & 1.8 & 1.8 & 1.15 & 1.15 \\
III& 2.2 & 2.2 & 1.05 & 1.05 \\
IV & 2.2 & 2.2 & 1.15 & 1.15 \\
\label{table1}
\end{tabular}
\end{center}
whereas  the other parameters values are: $v_{M_1}=v_{M_2}=1$,$K_{M_1}=K_{M_4}=1$ and $N_T=40$.
For the first parameter set, the zero-order approximation is a bimodal distribution, but the effect of the currents induced by the rotor potential $H$ (see fig. \ref{rot_fig})
left) are able to destroy the bimodal behaviour by depressing the two maximal at the border (see fig. 5).

If we increase the value of the $K_{2M}$ and $K_{3M}$ (case II), the exact stationary distribution becomes bimodal and the effect of currents is
to introduce a strong transition probability between the two distribution maxima (fig. \ref{dist02_fig}).

Therefore, when the MM constants $K_{M2}$ and $K_{M3}$ are $<2$ (we note that for $K_{M_2}=K_{M_3}=2$, the detailed balance holds), the non-conservative nature of the field (\ref{field_nc}) introduces
a delay in the phase transition from a mono-modal to a bimodal distribution. However, when we consider $K_{M2}=K_{M3}>2$ (cases III and IV)
the rotor potential $H$ moves the particle towards the borders and the central part of the distribution is depressed. This is shown in
the figure \ref{dist03_fig} where we compare the zero-order approximation of the stationary distribution and the solution of the CME
using the case III parameters of the table \ref{table1}.

Finally, in case IV of table \ref{table1}, the CME stationary solution undergoes a
transition to a bimodal distribution, whereas the zero-order approximation is still mono-modal ( fig. \ref{dist04_fig}), so that the
effect of currents is to anticipate the phase transition.

\begin{figure}[th]
\centerline{
\includegraphics[width=0.4\textwidth]{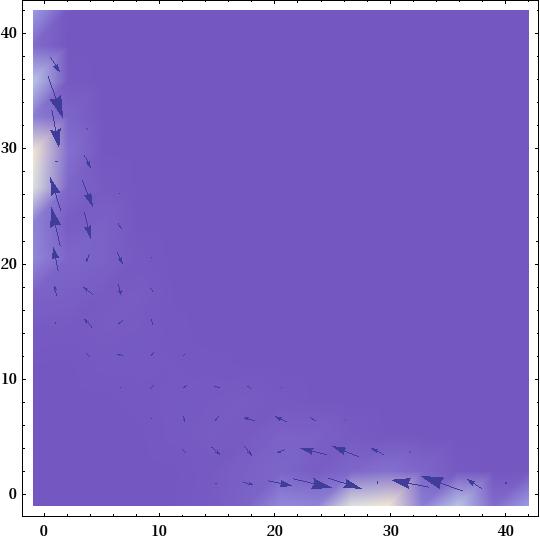}
}
\caption[]{ Current vector field computed by using definiton (\ref{current}) and the stationary solution of the CME with
case IV parameters. The distribution is bimodal (cfr. figure 8) and the current lines tend to be orthogonal to
the distribution gradient near the maximal value.}
\label{rotdist04_fig}
\end{figure}

\begin{figure*}[th]
\centerline{
\includegraphics[width=0.45\textwidth]{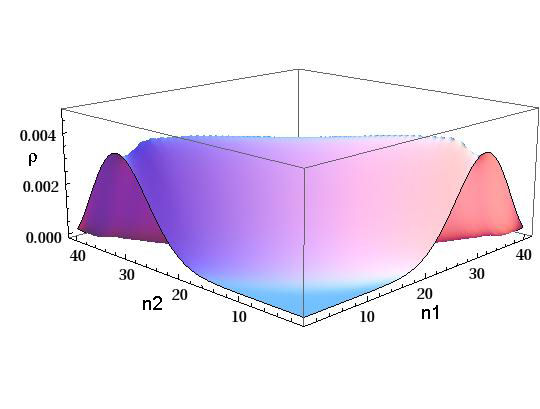}
\qquad
\includegraphics[width=0.45\textwidth]{fig_9a.jpg}
}
\caption[]{The same as in fig. \ref{dist01_fig} using parameter values of case IV in table \ref{table1}.}
\label{dist04_fig}
\end{figure*}
Using the stationary solution one can also compute the currents according to definition (\ref{current}). In figure (9)  we plot
the current vector in case IV parameters to show that the current tends to become normal to the distribution gradient near the
maximal value.

This result can be also understood using the perturbative approach (\ref{pot_pert}), where one shows that the main effect of the
$V_1$ potential correction is to compensate the rotor field of $H$ along the distribution gradient directions. As a consequence, the current is
zero at the maximal distribution value and condition (\ref{crit_point}) defines the critical points of the stationary distribution
even in the non-conservative case. 
\section{Conclusions}
The CME approach we present here is a powerful method for studying complex cellular processes, even with
significant simplifications such as spatial homogeneity of volumes where the chemical reactions are taking place.
The CME theory is attractive for a variety of reasons, including the richness of aspects
(the capability of coping with fluctuations and chemical fluxes) and the possibility of developing thermodynamics,
starting from the distribution function. The violation of detailed balance gives information
on the ``openness'' of the system and on the nature of the bistable regimes, which are induced by the external environment;
in contrast, it is a free-energy equilibrium when detailed balance holds. This statement can be expressed
in a more rigorous form by introducing the vector field generated from the ratio between the generation and recombination
terms, by decomposing it into a sum of ``conservative'' and ``rotational'' fields (Helmholtz decomposition)
and by relating the chemical fluxes to the non-conservative field. The magnitude of deviations from detailed
balance influences the form of the stationary distribution at the transition to a bistable regime, which may be
driven by the currents.
An interesting test for the prediction of the PdPC CME  model would be to perform one experiment with the parameter
values chosen to satisfy DB and compare it with another set of parameters where DB is not fulfilled.
Our results show that the PdPC can operate across these two regions, and that the transition regime
can be explained by the role of the currents, that, within a thermodynamic framework, can be interpreted
as the effect of an external energy source.
A full thermodynamic analysis of this cycle is beyond the scope of this paper, but we can surmise that this approach
might be extended to other cycles in order to quantify if, and how much energy, is required to maintain or create bistability.
Another way to extend this analysis would be the generalization to $n$-step phospho/dephosphorylation cycles,
where the stationary distribution will be the product of $n$ independent one-dimensional distributions.
In conclusion, our results could be important for a deeper characterization of biochemical signaling
cycles that are the molecular basis for complex cellular behaviors implemented as a ``switch'' between states.

\section{Appendix}

The master equation describes the evolution of one-step Markov Processes according to
\begin{eqnarray}
&&{\partial\rho\over \partial t}(n_1,n_2,t)=\nonumber\\
&&g_1(n_1-1,n_2)\rho(n_1-1,n_2,t)-g_1(n_1,n_2)\rho(n_1,n_2,t)\nonumber\\
&&+r_1(n_1+1,n_2)\rho(n_1+1,n_2,t)-r_1(n_1,n_2)\rho(n_1,n_2,t)\nonumber\\
&&+g_2(n_1,n_2-1)\rho(n_1,n_2-1,t)-g_2(n_1,n_2)\rho(n_1,n_2,t)\nonumber\\
&&+r_2(n_1,n_2+1)\rho(n_1,n_2+1,t)-r_2(n_1,n_2)\rho(n_1,n_2,t)\nonumber\\
\label{markov1app}
\end{eqnarray}
with the boundary conditions for the coefficients
\begin{eqnarray}
&&g_1(n,N-n)=g_2(n,N-n)=0 \quad  {\rm and} \quad \nonumber\\
&&r_2(n,0)=r_1(0,n)=0\qquad \; n\in[0,N_T]\nonumber\\
\label{coefcond}
\end{eqnarray}
so that $n_1+n_2\le N_T$. By introducing the difference operators (\ref{diff_op}),
eq. (\ref{markov1app}) can be written in the form of a continuity equation
\begin{equation}
{\partial\rho\over \partial t}(n_1,n_2,t)= -D_1  J_1(n_1,n_2,t) -D_2 J_2(n_1,n_2,t)
\label{markov1s_c}
\end{equation}
where we introduce the current vector $J$ of components:
\begin{eqnarray}
J_1(n_1,n_2,t)= g_1(n_1-1,n_2)\rho(n_1-1,n_2,t)\nonumber \\
-r_1(n_1,n_2)\rho(n_1,n_2,t)\nonumber \\
J_2(n_1,n_2,t)= g_2(n_1,n_2-1)\rho(n_1,n_2-1,t)\nonumber \\
-r_2(n_1,n_2)\rho(n_1,n_2,t)\nonumber \\
\end{eqnarray}
The stationary solution $\rho_s(n_1,n_2)$ is characterized by the zero divergence condition for the current
(25).
Detailed balance holds when the current is zero and $\rho_s$ satisfies
\begin{eqnarray}
&&r_1(n_1,n_2)\rho_s(n_1-1,n_2)\nonumber \\
\cdot&&\left (
{\rho_s(n_1,n_2)\over \rho_s(n_1-1,n_2)}-{g_1(n_1-1,n_2)\over r_1(n_1,n_2)}\right )=0\nonumber \\
&&r_2(n_1,n_2)\rho_s(n_1,n_2-1)\nonumber \\
\cdot&&\left (
{\rho_s(n_1,n_2)\over \rho_s(n_1,n_2-1)}-{g_2(n_1,n_2-1)\over r_2(n_1,n_2)}\right )=0\nonumber \\
\label{det_bal}
\end{eqnarray}
for $0<n_1$ and $0<n_2$. The previous equations can be written in the form
\begin{eqnarray}
D_1\ln(\rho_s(n_1-1,n_2))&=&\ln\left ({g_1(n_1-1,n_2)\over r_1(n_1,n_2)}\right )\nonumber \\
D_2\ln(\rho_s(n_1,n_2-1))&=&\ln\left ({g_2(n_1,n_2-1)\over r_2(n_1,n_2)}\right )\nonumber \\
\label{det_bal0}
\end{eqnarray}
and, if one introduces the the vector field 
\begin{eqnarray}
a_1(n_1,n_2)&=&{g_1(n_1,n_2)\over r_1(n_1+1,n_2)}\nonumber \\
a_2(n_1,n_2)&=&{g_2(n_1,n_2)\over r_2(n_1,n_2+1)}\nonumber \\
\label{det_bal00}
\end{eqnarray}
due to the commutative property of the difference operators $D_i$,
detailed balance implies an irrotational character for the vector field $\ln(a(n_1,n_2))$
\begin{equation}
D_2\ln (a_1(n_1,n_2))-D_1\ln (a_2(n_1,n_2))=0
\label{det_bal1}
\end{equation}
If we have no singularities in the domain, eq. (\ref{det_bal1}) is a sufficient condition for the existence of a potential $V(n_1,n_2)$
(cfr. eq, (\ref{det_bal2})) and
the distribution $\rho_s(n_1,n_2)$ can be computed using the recurrence relations
\begin{eqnarray}
\rho_s(n_1+1,n_2)&=& a_1(n_1,n_2)\rho_s(n_1,n_2)\nonumber \\
\rho_s(n_1,n_2+1)&=& a_2(n_1,n_2)\rho_s(n_1,n_2)\nonumber \\
\label{staz_db}
\end{eqnarray}
Therefore, the components $a_1(n_1,n_2)$ and $a_2(n_1,n_2)$ can also be interpreted as creation operators according to
relations (\ref{staz_db}) and detailed balance condition (\ref{det_bal1}) is equivalent to the commutativity property for these operators.
The stationary distribution can be written in the Maxwell-Boltzmann form (\ref{max_boltz}) and the potential $V(n_1,n_2)$ is
associated with an ``energy function".
We finally observe that the critical points of the stationary distribution $\rho_s$ are defined by the condition
\begin{equation}
a_1(n_1,n_2)=a_2(n_1,n_2)=1
\label{crit_point1}
\end{equation}
For the double phosphorylation cycle (\ref{double_cycle_eq_two_enz}) it is possible to derive explicit expressions
for the stationary distribution $\rho_s(n_1,n_2)$ by applying recursively the relations (\ref{staz_db})
in a specific order: for example, first moving along the $n_2$ direction and then along $n_1$, we obtain an expression for $\rho_s(n_1,n_2)$ as a function of
$\rho_s(0,0)$
\begin{equation}
\displaystyle{\rho_s(n_1,n_2)=\prod_{i=1}^{n_1}\frac{g_1(i-1,n_2)}{r_1(i,n_2)}\prod_{l=1}^{n_2}\frac{g_2(0,l-1)}{r_2(0,l)}\rho_s(0,0)}
\label{staz_dist_2D}
\end{equation}
A direct substitution of the coefficients (\ref{twoenz}) in the relation (\ref{staz_dist_2D}) gives
\begin{widetext}
\begin{eqnarray*}
&&\rho_s(n_1,n_2)=\prod_{i=1}^{n_1}\frac{K_{M4}V_{M2}(N_T-i-n_2+1)}{K_{M2}K_{M4} + K_{M4}(N_T-i-n_2+1)+K_{M2}n_2}\cdot
\frac{K_{M1}K_{M3}+K_{M1}(N_T-i-n_2) + K_{M3}i}{K_{M3}V_{M1}i}\nonumber \\
&&\cdot \prod_{l=1}^{n_2}\frac{K_{M1}V_{M3}(N_T-l+1)}{K_{M1}K_{M3} + K_{M1}(N_T-l+1)}
\frac{K_{M2}K_{M4} + K_{M4}(N_T-l)+K_{M2} l}{K_{M2}V_{M4} l}  \rho_s(00)\nonumber \\
\end{eqnarray*}
\end{widetext}

We can further simplify this expression by using the definition of multinomial coefficients and the rising and falling factorial symbols, defined as $x^{(n)}=x(x+1)(x+2)\cdots(x+n-1)= \frac{(x+n-1)!}{(x-1)!}$ and $x_{(n)}=x(x-1)(x-2)\cdots(x-n+1)=\frac{x!}{(x-n)!}$) respectively.

\begin{eqnarray*}
&&\rho_s(n_1,n_2)=\left(\frac{V_{M2}}{V_{M1}}\right)^{n_1}\left(\frac{V_{M3}}{V_{M4}}\right)^{n_2}{N_T-n_2 \choose n_1} 
{N_T \choose n_2}\nonumber\\
&&\left(\frac{K_{M3}-K_{M1}}{K_{M3}}\right)^{n_1}
\left(\frac{K_{M2}-K_{M4}}{K_{M2}}\right)^{n_2}\cdot\nonumber\\
&&\frac{(K_{M1}(1+n_2-K_{M3}-N_T)-K_{M3})^{(n_1)}}{(K_{M1}-K_{M3})^{(n_1)}(K_{M2}(1+\frac{n_2}{K_{M4}})+N_T-n_2)_{(n_1)}}\cdot\nonumber\\
&&\frac{(K_{M2}(1+K_{M4})+K_{M4}(N_T-1))^{(n_2)}}{(K_{M2}-K_{M4})^{(n_2)}(K_{M3}+N_T)_{(n_2)}}\rho_s(0,0)\nonumber\\
\end{eqnarray*}


Finally, it is interesting to go to a continuous limit that is equivalent to $N_T\to \infty$. First we introduce the population
densities $A=n_1/N_T$ and $A_P^P=n_2/N_T$ and use the fact that the generation and recombination
rates are invariant by substituting $n_1$ and $n_2$ with $A$ and $A_P^P$. Then we approximate
\begin{eqnarray*}
D_1 V(A,A_P^P)=V(A+1/N_T,A_P^P)-V(A,A_P^P)\nonumber \\
={1\over N_T}{\partial V\over \partial A}(A+{1\over 2N_T},A_P^P)+O(1/N_T^3)\nonumber \\
\end{eqnarray*}
and a similar expression holds for $D_2 V(A,A_P^P)$. According to eq. (\ref{det_bal2}), the partial derivatives of $V(A,A_P^P)$
are bounded when $N_T\to \infty$ only in the domain where the following approximation holds (diffusion dominated
region)
\begin{equation}
{g_i(A,A_P^P)\over r_i(A,A_P^P)}=1+O(1/N_T) \qquad i=1,2
\label{cont_lim}
\end{equation}
and we can estimate
\begin{eqnarray}
&&\ln\left ({g_1(A,A_P^P)\over r_1(A+1/N_T,A_P^P)}\right )\nonumber \\
&\simeq& -2{r_1(A+1/N_T,A_P^P)-g_1(A,A_P^P)\over r_1(A+1/N_T,A_P^P)+g_1(A,A_P^P)}
+O(1/N_T^3)\nonumber \\
&&\ln\left ({g_2(A,A_P^P)\over r_2(A,A_P^P+1/N_T)}\right )\nonumber \\
&\simeq& -2{r_2(A,A_P^P+1/N_T)-g_2(A,A_P^P)\over r_2(A,A_P^P+1/N_T)+g_2(A,A_P^P)}+
O(1/N_T^3)\nonumber \\
\end{eqnarray}
Then we may approximate (we use the convention of leaving out the dependence on $A_P^P$)
\begin{widetext}
\begin{eqnarray}
{r_1(A+1/N_T)-g_1(A)\over r_1(A+1/N_T)+g_1(A)}=
{r_1(A+1/2N_T)-g_1(A+1/2N_T)+{1/2N_T}\left ({\partial r_1/\partial A}(A+1/2N_T)
+{\partial g_1/\partial A}(A+1/2N_T)\right )
\over r_1(A+1/N_T)+g_1(A)}\nonumber \\
\end{eqnarray}
\end{widetext}
up to an error of order $O(1/N_T^2)$ (a similar expression is obtained for the second equation).
Therefore, detailed balance in the continuous limit reads
\begin{widetext}
\begin{eqnarray}
{r_1(A,A_P^P)-g_1(A,A_P^P)+{1/2N_T}\left ({\partial r_1/\partial A}(A,A_P^P)
+{\partial g_1/\partial A}(A,A_P^P)\right )\over r_1(A+1/2N_T,A_P^P)+g_1(A-1/2N_T,A_P^P)}=
-{1\over 2N_T}{\partial V\over \partial A}(A,A_P^P)\nonumber \\
{r_2(A,A_P^P)-g_2(A,A_P^P)+{1/2N_T}\left ({\partial r_2/\partial A_P^P}(A,A_P^P)
+{\partial g_2/\partial A_P^P}(A,A_P^P)\right )\over r_2(A,A_P^P+1/2N_T)+g_2(A,A_P^P-1/2N_T)}=
-{1\over 2N_T}{\partial V\over \partial A_P^P}(A,A_P^P)\nonumber \\
\label{det_balc}
\end{eqnarray}
\end{widetext}
The limit $N_T\to \infty$ turns out to be singular since

\begin{widetext}
\begin{eqnarray}
-{\partial V\over \partial A}(A,A_P^P)&=&{ 2N_T(r_1(A,A_P^P)-g_1(A,A_P^P))+{\partial r_1/\partial A}(A,A_P^P)
+{\partial g_1/\partial A}(A,A_P^P)\over r_1(A,A_P^P)+g_1(A,A_P^P)}\nonumber \\
-{\partial V\over \partial A_P^P}(A,A_P^P)&=&{ 2N_T(r_2(A,A_P^P)-g_2(A,A_P^P))+{\partial r_2/\partial A_P^P}(A,A_P^P)
+{\partial g_2/\partial A_P^P}(A,A_P^P)\over r_2(A,A_P^P)+g_2(A,A_P^P)}\nonumber \\
\label{det_balt}
\end{eqnarray}
\end{widetext}
Hence in the diffusion domain defined by condition (\ref{cont_lim}), we recover detailed balance
for a Fokker-Planck equation
with drift and diffusion coefficients defined as:
$$
c_i(A,A_P^P)=2N_T\left (r_i(A,A_P^P)-g_i(A,A_P^P)\right )\qquad i=1,2
$$
and
$$
b_i(A,A_P^P)=r_i(A,A_P^P)+g_i(A,A_P^P)\qquad i=1,2
$$
In the diffusion region the drift and the diffusion coefficients are of the same order, otherwise
$c_i(A,A_P^P)\gg b_i(A,A_P^P)$ when $N_T\gg 1$.
As a consequence, the stationary solution of F.P. equation is an approximation of the stationary distribution of the CME in the diffusion region,
but the approximation of the transient state dynamics using the F.P. equation requires further studies due to the singularity of the
thermodynamic limit.
\par\noindent
In the generic case of eq. (\ref{markov1app}), we represent the r.h.s. of eq. (\ref{det_bal0})
as a sum of a rotational and a gradient vector fields
\begin{eqnarray}
\ln a_1(n_1,n_2)&=&D_1 V(n_1,n_2)+D_2 H(n_1,n_2)\nonumber \\
\ln a_2(n_1,n_2)&=&D_2 V(n_1,n_2)-D_1 H(n_1,n_2)\nonumber \\
\label{campo}
\end{eqnarray}
Taking into account the condition $n_1+n_2\le N_T-1$, from the eqs. (\ref{campo}) we get
the discrete Poisson equations
\begin{eqnarray}
(D_1D_1 +D_2D_2) V(n_1,n_2)=\nonumber \\
D_1\ln\left (a_1(n_1,n_2)\right )+
D_2\ln\left (a_2(n_1,n_2)\right )\nonumber \\
(D_1 D_1 +D_2D_2) H(n_1,n_2)=\nonumber \\
D_2\ln\left (a_1(n_1,n_2)\right )-
D_1\ln\left (a_2(n_1,n_2)\right )\nonumber \\
\label{campo1}
\end{eqnarray}
We remark that the r.h.s. of eqs. (\ref{campo1}) is defined only if $n_1+n_2\le N-2$ and corresponds to
$N(N-1)/2$ independent equations, whereas we have $(N+2)(N+1)/2$ unknown values $H(n_1,n_2)$.
As a consequence from the explicit form of the discrete Poisson operator
\begin{widetext}
$$
(D_1^2 +D_2^2) H(n_1,n_2)=H(n_1+2,n_2)-2H(n_1+1,n_2)+H(n_1,n_2)+H(n_1,n_2+2)-2H(n_1,n_2+1)+H(n_1,n_2)
$$
\end{widetext}
we can set the boundary conditions $H(n,N-n)=H(n,N-1-n)=0$ and recursively solve the system setting
\begin{eqnarray}
2H(n,N-2-n)=\nonumber \\
\ln\left ( {a_1(n,N-1-n) a_2(n,N-2-n)\over a_1(n,N-2-n) a_2(n+1,N-2-n)}
\right ))\nonumber \\
\end{eqnarray}
and successively using the equations
\begin{widetext}
\begin{eqnarray}
2H(n,N-j-n)=-H(n+2,n-j-n)-H(n,N-j-n-2)
+2H(n+1,N-j-n)+2H(n,N-j-n-1)\nonumber \\+\ln {a_1(n,N-j-n+1) a_2(n,N-j-n)\over a_1(n,N-j-n) a_2(n+1,N-j-n)}\nonumber \\
\label{recurs}
\end{eqnarray}
\end{widetext}
for $N\ge j>2$. Once $H(n_1,n_2)$ is computed, we define the ``potential" $V(n_1,n_2)$ by using eq.
(\ref{campo}). The recursion relations (\ref{recurs}) can be written in an exponential form by defining
$$
R(n_1,n_2)=\exp (H(n_1,n_2))
$$
\begin{eqnarray}
R(n,N-j-n))=\nonumber \\
\sqrt{{a_1(n,N-j-n+1) a_2(n,N-j-n)\over a_1(n,N-j-n) a_2(n+1,N-j-n)}}\nonumber \\
\cdot{ R(n+1,N-j-n)R(n,N-j-n-1)\over \sqrt{R(n+2,n-j-n)R(n,N-j-n-2)}}\nonumber \\
\end{eqnarray}
As a consequence, the recurrence (\ref{staz_db}) reads
\begin{eqnarray}
\rho_s(n_1+1,n_2)&=& a_1(n_1,n_2){R(n_1,n_2)\over R(n_1,n_2+1)}\rho_s(n_1,n_2)\nonumber \\
\rho_s(n_1,n_2+1)&=& a_2(n_1,n_2){R(n_1+1,n_2)\over R(n_1,n_2)}\rho_s(n_1,n_2)\nonumber \\
\end{eqnarray}
for all $n_1+n_2\le N-1$.\par\noindent
The current componentss (25) turn out to be proportional to the rotational part of the field
(\ref{campo}) (i.e. to $H(n_1,n_2)$)\cite{Onsag31}, so that the current vanishes at the points where condition (\ref{crit_point1}) is satisfied.
One can prove that the critical points of the stationary distribution are still defined by eq. (\ref{crit_point1}).
Indeed, if one computes the formal expansion of the generation and recombination rates around a solution of eqs. (31)
\begin{eqnarray}
g(n)&=&1+{\partial g\over \partial n}(n^\ast)\cdot\Delta n +....\nonumber \\
r(n)&=&1+{\partial r\over \partial n}(n^\ast)\cdot\Delta n +....\nonumber \\
\end{eqnarray}
(for the sake of simplicity we have normalized the value of the generation and recombination rate to 1 at the critical point)
the current components can be approximated by the expressions
\begin{eqnarray}
J_1^s(n_1,n_2)\simeq \rho_s(n_1-1,n_2)-\rho_s(n_1,n_2)\nonumber \\+{\partial g_1\over \partial n}(n^\ast)\cdot\Delta n\rho_s(n_1-1,n_2)-
{\partial r_1\over \partial n}(n^\ast)\cdot\Delta n\rho_s(n_1,n_2)\nonumber \\
J_2^s(n_1,n_2)\simeq \rho_s(n_1,n_2-1)-\rho_s(n_1,n_2)\nonumber \\+{\partial g_2\over \partial n}(n^\ast)\cdot\Delta n\rho_s(n_1,n_2-1)-
{\partial r_2\over \partial n}(n^\ast)\cdot\Delta n\rho_s(n_1,n_2)\nonumber \\
\label{current_app}
\end{eqnarray}
At the critical point $n_\ast$ we get
\begin{equation}
\rho_s(n_1^\ast-1,n_2^\ast)=\rho_s(n_1^\ast,n_2^\ast-1)=\rho_s(n_1^\ast,n_2^\ast)
\label{distri_crit}
\end{equation}
since $J^s(n_1^\ast,n_2^\ast)=0$. The condition (\ref{distri_crit}) means the $n^\ast$ is a critical point for the stationary
distribution $\rho_s$.
\par\noindent
When $H(n_1,n_2)$ is small, the detailed balance solution (\ref{max_boltz}) is a good approximation of the stationary
solution $\rho_s(n_1,n_2)$ and a perturbative approach can be applied.
Let us write the stationary condition (\ref{staz_sol}) in the form
\begin{eqnarray}
D_1 r_1(n_1,n_2)\rho_s(n_1,n_2)\left ( 1-a_1(n_1-1,n_2){\rho_s(n_1-1,n_2)\over \rho_s(n_1,n_2) }\right )&+&\nonumber \\
D_2 r_2(n_1,n_2)\rho_s(n_1,n_2)\left ( 1-a_2(n_1,n_2-1){\rho_s(n_1,n_2-1)\over \rho_s(n_1,n_2) }\right )&=&0\nonumber \\
\end{eqnarray}
By using the definitions (\ref{campo}), we assume that the rotational field is associated with a potential $\epsilon H(n_1,n_2)$, with $\epsilon \ll 1$
perturbation parameter and we write the stationary solution in the form
\begin{equation}
\rho_s(n_1,n_2)=C\exp(V(n_1,n_2)+\epsilon V_1(n_1,n_2))
\label{staz_pert}
\end{equation}
From a direct calculation we get
\begin{widetext}
\begin{eqnarray}
D_1 r_1(n_1,n_2)e^{V(n_1,n_2)}\left ( 1-\exp(\epsilon (D_2 H(n_1-1,n_2)-D_1V_1(n_1-1,n_2))) \right )\nonumber \\
+D_2 r_2(n_1,n_2)e^{V(n_1,n_2)}\left ( 1-\exp(-(\epsilon D_1 H(n_1,n_2-1)+D_2 V_1(n_1,n_2-1))\right )\simeq\nonumber \\
\epsilon D_1 r_1(n_1,n_2)e^{V(n_1,n_2)}\left ( D_1 V_1(n_1-1,n_2)-D_2 H(n_1-1,n_2) \right )\nonumber \\
+\epsilon D_2 r_2(n_1,n_2)e^{V(n_1,n_2)}\left ( D_2V_1(n_1,n_2-1)+D_1 H(n_1,n_2-1) \right )=0\nonumber \\
\label{staz_eq2}
\end{eqnarray}
\end{widetext}
for all the values $n_1+n_2\le N-1$ and $n_i\ge 1$.
The correction potential $V_1(n_1,n_2)$ has to be computed from the previous equation, and it enters in the
definition of the stationary currents.\par\noindent

\end{document}